\documentstyle[prb,aps,eqsecnum,preprint,tighten,epsf,floats]{revtex}
\begin{document}
\draft
\title{Thermopower of Single-Channel Disordered and Chaotic Conductors}
\author{S. A. van Langen$^1$, P. G. Silvestrov$^{1,2}$, and C. W. J. Beenakker$^1$}
\address{$^1$ Instituut-Lorentz, Leiden University,
P.O. Box 9506, 2300 RA Leiden, The Netherlands \\
$^2$ Budker Institute of Nuclear Physics, Novosibirsk, Russia}
\maketitle
\begin{abstract}
We show (analytically and by numerical simulation) that the
zero-temperature limit of the distribution of the thermopower $S$ of a
one-dimensional disordered wire in the localized regime is a
Lorentzian, with a disorder-independent width of $4\pi^{3}k_{\rm
B}^2T/3e\Delta$ (where $T$ is the temperature and  $\Delta$ the mean level
spacing). Upon raising the temperature the distribution crosses over to
an exponential form $\propto\exp\left(-2|S|eT/\Delta\right)$.
We also consider the case of a chaotic quantum dot with two
single-channel ballistic point contacts. The distribution of $S$ then
has a cusp at $S=0$ and a tail $\propto |S|^{-1-\beta}\ln |S|$ for
large $S$ (with $\beta=1,2$ depending on the presence or absence of
time-reversal symmetry).
\end{abstract}

\section{Introduction}
Thermo-electric transport properties of conductors probe the energy
dependence of the scattering processes limiting conduction.  At low
temperatures and in small (mesoscopic) systems, elastic impurity
scattering is the dominant scattering process. The energy dependence of
the conductance is then a quantum interference effect.\cite{imry} The
derivative $dG/dE$ of the conductance with respect to the Fermi energy
is measured by the thermopower $S$, defined as the ratio $-\Delta
V/\Delta T$ of a (small) voltage and temperature difference applied
over the sample at zero electrical current.  Experimental and
theoretical studies of the thermopower exist for several mesoscopic
devices. One finds a series of sharp peaks in the thermopower of
quantum point contacts,\cite{qpc} aperiodic fluctuations in diffusive
conductors,\cite{diffusive} sawtooth oscillations in quantum dots in
the Coulomb blockade regime,\cite{beenstar} and Aharonov-Bohm
oscillations in metal rings.\cite{blanter}

Here we study the statistical distribution of the thermopower in two
different systems, not considered previously: A disordered wire in the
localized regime and a chaotic quantum dot with ballistic point
contacts. A single transmitted mode is assumed in both cases. In the
disordered wire, conduction takes place by resonant tunneling through
localized states. The resonances are very narrow and appear at
uncorrelated energies.  The distributions of the thermopower and the
conductance are both broad, but otherwise quite different: Instead of
the log-normal distribution of the conductance\cite{imry} we find a
Lorentzian distribution for the thermopower. In the quantum dot, the
resonances are correlated and the widths are of the same order as the
spacings. The correlations are described by random-matrix
theory,\cite{beenakker,weidenmuller} under the assumption that the
classical dynamics in the dot is chaotic. The thermopower distribution
in this case follows from the distribution of the time-delay
matrix found recently.\cite{brouwer}

The thermopower (at temperature $T$ and Fermi energy $E_{\rm F}$) is
given by the Cutler-Mott formula\cite{cutler,sivan}
\begin{equation}
\label{cutler}
S=-\frac{1}{eT}\frac{\int dE\,(E-E_{\rm F})G(E)df/dE}
{\int dE\,G(E)df/dE},\label{cm}
\end{equation}
where $G$ is the zero-temperature conductance and $f$ is the
Fermi-Dirac distribution function. In the limit $T\rightarrow 0$
Eq.\ (\ref{cm}) simplifies to 
\begin{equation}
S=-\frac{\pi^2}{3}\frac{k_{\rm B}^2T}{eG}\frac{dG}{dE}, 
\end{equation}
where $G$ and $dG/dE$ are to be evaluated at $E=E_{\rm F}$.  We
consider mainly the zero-temperature limit of the thermopower, by
studying the dimensionless quantity
\begin{equation}
\label{defsigma}
\sigma=\frac{\Delta}{2\pi G}\frac{dG}{dE}.
\end{equation}
Here $\Delta$ is the mean level spacing near the Fermi energy. Since we
are dealing with single-channel conduction, the conductance is related
to the transmission probability $T(E)$ by the Landauer formula\cite{imry,landauer}
\begin{equation}
\label{landauer}
G(E)=\frac{2e^2}{h}T(E).
\end{equation}
The problem of the distribution of the thermopower is therefore a problem of
the distribution of the logarithmic derivative of the transmission probability.

\section{Disordered wire}
In this section we study a disordered single-mode wire of length $L$
much greater than the mean free path $l$. This is the localized regime.
We compute the thermopower distribution in the zero-temperature limit.
The analytical theory is tested by comparing with a
numerical simulation. The effect of a finite temperature is
considered at the end of the section. Electron-electron interactions
play an important role in one-dimensional conduction, but we do not
take these into account here. 

\subsection{Analytical theory}
The localization length $\xi (E)$ [which is of order $l$ and is defined
by $\lim_{L\rightarrow\infty} L^{-1}\ln T(E)=-2/\xi (E)$] and the
density of states $\rho (E)$ [per unit of length in the limit
$L\rightarrow\infty$] are related by the Herbert-Jones-Thouless
formula\cite{hjt}
\begin{equation}
\label{hjt}
\frac{1}{\xi (E)} = \int dE' \rho (E') \ln |E-E'| + {\rm constant}.
\end{equation}
The additive constant is energy-independent on the scale of the level
spacing.  Eq.\ (\ref{hjt}) follows from the Kramers-Kronig relation
between the real and imaginary parts of the wave number (the real part
determining $\rho$, the imaginary part $\xi$).  Neglecting the width of
the resonances in the large-$L$ limit, the density of states $\rho(E)
=L^{-1}\sum_i\delta (E-E_i)$ is a sum of delta functions, and thus
\begin{equation}
\sigma 
=-\frac{L\Delta}{\pi}\frac{d}{dE}\frac{1}{\xi (E)}
=\frac{\Delta}{\pi}\sum_i\frac{1}{E_i-E_{\rm F}}.
\end{equation}

In the localized regime the energy levels $E_i$ are uncorrelated, and we
assume that they are uniformly distributed in a band of width $B$ around
$E_{\rm F}$. To obtain the distribution of $\sigma$,
\begin{equation}
P(\sigma )= \prod_i\int_{-B/2}^{B/2} \frac{dE_i}{B}\,\delta\left(
\sigma-\frac{\Delta}{\pi}\sum_j\frac{1}{E_j}\right),
\end{equation}
we first compute the Fourier transform
\begin{eqnarray}
P(k)=\int_{-\infty}^{\infty}d\sigma\,{\rm e}^{{\rm i}k\sigma} P(\sigma) =
\left[\frac{1}{B}\int_{-B/2}^{B/2}dE\,{\rm e}^{{\rm i}k\Delta/\pi
E}\right]^{B/\Delta} = {\rm e}^{-|k|},
\end{eqnarray}
where the limit $B/\Delta\rightarrow\infty$ is taken in the last step.
Inverting the Fourier transform, we find that the thermopower distribution is a
Lorentzian,
\begin{equation}
\label{lorentzian}
P(\sigma)=\frac{1/\pi}{1+\sigma^2}.
\end{equation}
The ``full width at half maximum'' of $P(\sigma)$ is equal to $2$, hence it is equal
to $4\pi^3k_{\rm B}^2T/3e\Delta$ for $P(S)$. This width depends on the length $L$ of the
system (through $\Delta\propto 1/L$), but it does not depend on the mean free path $l$
(as long as $l\ll L$, so that the system remains in the localized regime).

\subsection{Numerical simulation}

\begin{figure}[t]
\epsfxsize=0.6\hsize
\hspace*{\fill}
\vspace*{-0ex}\epsffile{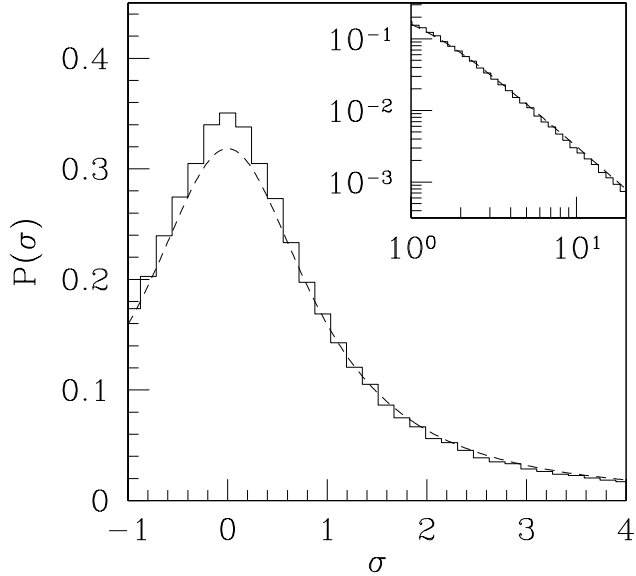}\vspace*{0ex}
\hspace*{\fill}
\medskip
\caption[]{
Distribution of the dimensionless thermopower $\sigma=(\Delta/2\pi)d\ln T(E)/dE$ for
a one-dimensional wire in the localized regime. The histogram is obtained
from a numerical simulation, for a sample length $L=32.3\,\xi$. The dashed
curve is the Lorentzian (\protect\ref{lorentzian}), being the
analytical result for $L\gg\xi$. The inset shows the algebraic tail of
the distribution on a logarithmic scale. The thermopower $S$, in the
zero-temperature limit, is related to $\sigma$ by
$S=-(2\pi^{3}/3)(k_{\rm B}^{2}T/e\Delta)\sigma$.}
\label{figwire}
\end{figure}

In order to check the analytical theory, we performed
a numerical simulation using the tight-binding Hamiltonian
\begin{equation}
{\cal H} = -{w \over 2} \sum_{j} \left(c^{\dagger}_{j+1}
c^{\vphantom{\dagger}}_{j} +  c^{\dagger}_{j}
c^{\vphantom{\dagger}}_{j+1} \right) + \sum_{j}
V_{j}^{\vphantom{\dagger}} c^{\dagger}_{j} c^{\vphantom{\dagger}}_{j}.
\label{eq:Hlattice}
\end{equation}
The disordered wire was modeled by a chain of lattice constant $a$,
with a random impurity potential $V_{j}$ at each site drawn from a
Gaussian distribution of mean zero and variance $u^2$. The localization
length of the wire is given by\cite{Dor88} $\xi=2(a/u^{2})(w^{2}-E_{\rm
F}^{2})$. We have chosen $u=0.075\,w$, $E_{\rm F}=-0.55\,w$, such that
$\xi=248\,a$, much smaller than $L=8000\,a$. From the scattering matrix
we obtained the conductance via the Landauer formula (\ref{landauer}),
and then the
(dimensionless) thermopower via Eq.\ (\ref{defsigma})
(with $\Delta =3.3\cdot 10^{-4}\,w$).
The differentiation with respect to energy was done numerically, by 
repeating the calculation at two closely spaced values of $E_{\rm F}$.
As shown in Fig.\ \ref{figwire}, the agreement with the analytical result is
good without any adjustable parameters.

\subsection{Finite temperatures}
\label{fintemp}

Our derivation of the Lorentzian distribution of the thermopower holds if the
temperature is so low
that $k_{\rm B}T$ is small compared to the typical width $\gamma$ of the transmission
resonances. What if $k_{\rm B}T>\gamma$, but still $k_{\rm B}T\ll\Delta$ (so that the
discreteness of the spectrum remains resolved)? We will show that the distribution
crosses over to an exponential, but in a highly non-uniform way.

Consider arbitrary $\gamma$ and $k_{\rm B}T$, both $\ll\Delta$. The 
Cutler-Mott formula (\ref{cutler}) is dominated by two contributions, one from a
peak in $df/dE$ of width $k_{\rm B}T$ around $E_{\rm F}$ and
one from a peak in $G(E)$ of width $\gamma_0$ around $E_0$. Here $\gamma_0$ and $E_0$
are the width and position of the level closest to $E_{\rm F}$. If
$|E_{\rm F}-E_0|\gg {\rm max}\, (k_{\rm B}T,\gamma_0)$, the two peaks do not overlap
and one can estimate the thermopower as
\begin{eqnarray}
\label{twopeak}
S &=& \frac{1}{eT} \left[ \frac{\pi\gamma_0 (k_{\rm B}T)^2}{3(E_{\rm F}-E_0)^3}+
\frac{E_{\rm F}-E_0}{k_{\rm B}T}
{\rm e}^{-|E_{\rm F}-E_0|/k_{\rm B}T} \right] \nonumber \\
&& \times \left[ \frac{\gamma_0}{2\pi(E_{\rm F}-E_0)^2} +\frac{1}{k_{\rm B}T}
{\rm e}^{-|E_{\rm F}-E_0|/k_{\rm B}T} \right] ^{-1}.
\end{eqnarray}
If $k_{\rm B}T\ll\gamma_0$, the first terms in numerator and denominator dominate
over the second terms. This is the regime that the Lorentzian distribution
(\ref{lorentzian}) holds for all $S$. 

We now turn to the regime $k_{\rm B}T>\gamma_0$. The first terms dominate
if $|E_{\rm F}-E_0|\gg k_{\rm B}T\ln k_{\rm B}T/\gamma_0$
Hence $P(S)$ is a Lorentzian
for $|S|\ll(k_{\rm B}/e)\left(\ln k_{\rm B}T/\gamma_0\right)^{-1}$.
The logarithm of $k_{\rm B}T/\gamma_0$ can be quite large, because the width of the
levels is exponentially
small in the system size, $\gamma \sim {\rm e}^{-L/\xi}$. The Lorentzian persists
in an interval larger than its width, provided
$k_{\rm B}T < \Delta \left(\ln k_{\rm B}T/\gamma_0 \right)^{-1}$. The second terms in
Eq.\ (\ref{twopeak}) dominate if
$k_{\rm B}T\ll |E_{\rm F}-E_0|\ll k_{\rm B}T\ln k_{\rm B}T/\gamma_0$.
In this case the thermopower is simply $S=(E_{\rm F}-E_0)/eT$, with exponential distribution
\begin{equation}
\label{exponential}
P(S) = \frac{eT}{\Delta} {\rm e}^{-2|S|eT/\Delta}.
\end{equation}
The distribution (\ref{exponential}) follows because the energy levels are uncorrelated,
so that the spacing $|E_{\rm F}-E_0|$ has an exponential distribution with a mean of
$\Delta/2$.

\begin{figure}[hb!]
\epsfxsize=0.6\hsize
\hspace*{\fill}
\vspace*{-0ex}\epsffile{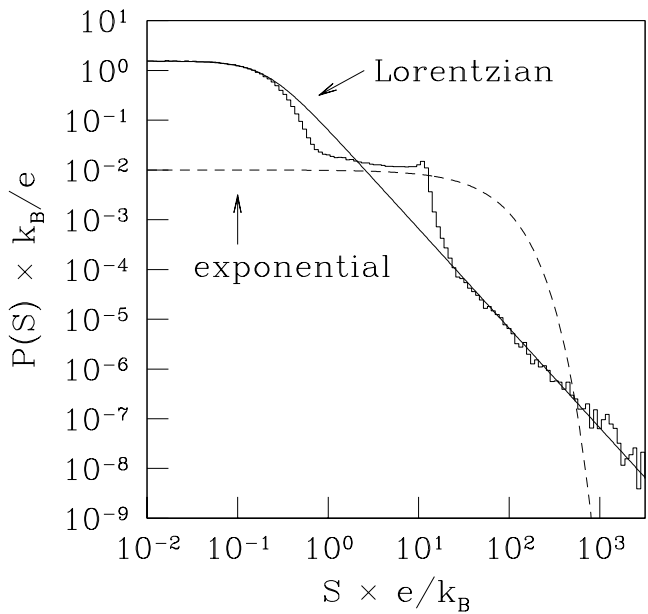}\vspace*{0ex}
\hspace*{\fill}
\medskip
\caption[]{ 
Thermopower distribution of a one-dimensional wire in the
localized regime at finite temperature. The histogram is obtained
from Eqs.\ (\ref{cutler}) and (\ref{Gfinitewidth}), by numerical integration for a set of
randomly chosen energy levels $E_i$, all having the same width
$\gamma_i=\gamma = 10^{-6}\Delta$. The temperature is $k_{\rm B}T/\Delta=0.01$,
such that $\gamma\ll k_{\rm B}T\ll\Delta$. The distribution follows the Lorentzian
(\ref{lorentzian}) (solid curve) for small and large $S$, but it follows the exponential
(\ref{exponential}) (dashed curve) in an intermediate region.}
\label{crossover}
\end{figure}

We conclude that the thermopower distribution for $\gamma < k_{\rm B}T\ll\Delta$
contains both Lorentzian and exponential contributions. The peak region
$|S|\ll (k_{\rm B}/e)\left(\ln k_{\rm B}T/\gamma\right)^{-1}$ is the Lorentzian
(\ref{lorentzian}). The intermediate region
$(k_{\rm B}/e)\left(\ln k_{\rm B}T/\gamma\right)^{-1}
\ll |S|\ll (k_{\rm B}/e)\ln k_{\rm B}T/\gamma$
is the exponential (\ref{exponential}). The far tails
$|S|\gg (k_{\rm B}/e)\ln k_{\rm B}T/\gamma$ can not be explained
by Eq.\ (\ref{twopeak}). With increasing temperature, the Lorentzian peak region
shrinks, and ultimately the exponential region starts right at $S=0$. This applies to
the temperature range
$\Delta \left(\ln k_{\rm B}T/\gamma \right)^{-1} < k_{\rm B}T \ll \Delta$.

To illustrate these various regimes, we computed $P(S)$ numerically from
Eq.\ (\ref{cutler}). We took the density of states 
\begin{equation}
\rho (E) = L^{-1}\sum_i \frac{\gamma_i/2\pi}{(E-E_i)^2 +\gamma_i^2/4}, 
\end{equation}
so that the conductance according to Eq.\ (\ref{hjt}) has the energy dependence
\begin{equation}
\label{Gfinitewidth}
G(E) \propto \prod_i \left[ (E-E_i)^2 + \gamma_i^2/4 \right]^{-1}.
\end{equation}
The levels $E_i$
were chosen uniformly and independently (mean spacing $\Delta$), but the fluctuations
of the widths $\gamma_i$ were ignored ($\gamma_i\equiv\gamma$ for all $i$).
Such fluctuations are irrelevant in the low-temperature
limit $k_{\rm B}T\ll\gamma$, but not for $\gamma<k_{\rm B}T\ll\Delta$. We believe
that ignoring fluctuations in $\gamma_i$ should still be a reasonable approximation,
because $\gamma_0$ appears only in logarithms.
The resulting $P(S)$ is plotted in Fig.\ \ref{crossover}.
We see the expected crossover from a Lorentzian to an exponential. The exponential region
appears as a plateau. Beyond the exponential region, the distribution appears to
return to the Lorentzian form. We have no explanation for this far tail.

\section{Chaotic quantum dot}
\begin{figure}
\epsfxsize=0.6\hsize
\hspace*{\fill}
\vspace*{-0ex}\epsffile{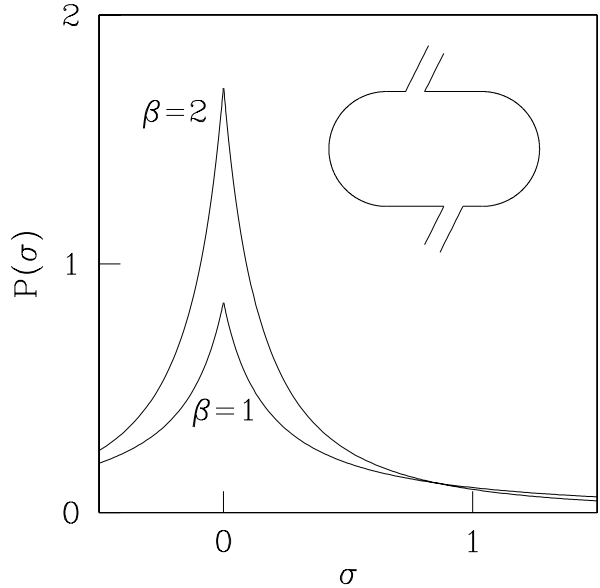}\vspace*{0ex}
\hspace*{\fill}
\medskip
\caption[]{
Distribution of the dimensionless thermopower of a
chaotic cavity with two single-channel ballistic point contacts (inset),
computed from Eq.\ (\protect\ref{tpdistcav}) for the case of broken ($\beta=2$)
and unbroken ($\beta=1$) time-reversal symmetry.}
\label{figcav}
\end{figure}

In this section we consider a chaotic quantum dot with single-channel
ballistic point contacts (see Fig.\ \ref{figcav}, inset). Because there
are no tunnel barriers in the point contacts, the effects of the
Coulomb blockade are small and here we ignore them altogether. For this
system, the distribution of $dT/dE$ was computed recently from
random-matrix theory.\cite{brouwer} The energy derivative of the
transmission probability has the parametrization
\begin{equation}
\frac{dT}{dE}=\frac{c}{\hbar}(\tau_1-\tau_2)\sqrt{T(1-T)},
\end{equation}
with independent distributions 
\begin{eqnarray}
&&P(c)\propto (1-c^2)^{-1+\beta /2},\;|c|<1, \\
&&P(\tau_1,\tau_2)\propto |\tau_1-\tau_2|^\beta (\tau_{1}\tau_{2})^{-2(\beta+1)}
{\rm e}^{-(1/\tau_1+1/\tau_2)\pi\beta\hbar/\Delta},\;\tau_{1},\tau_{2}>0,\\
&&P(T) \propto T^{-1+\beta/2},\;0<T<1.
\end{eqnarray}
The integer $\beta$ equals 1 or 2, depending on whether time-reversal
symmetry is present or not. The times $\tau_1,\tau_2$ are the
eigenvalues of the Wigner-Smith time-delay matrix
(see Refs.\ \onlinecite{brouwer} and \onlinecite{fyodorov}). Their sum
$\tau_1+\tau_2$ is the density of states (multiplied by $2\pi\hbar$).
The thermopower distribution follows from
\begin{eqnarray}
\label{tpdistcav}
P(\sigma)&\propto&\int_{-1}^1 dc\,P(c)\int_0^\infty d\tau_1\int_0^\infty
d\tau_2\,P(\tau_1,\tau_2)\int_0^1 dT\,P(T)\nonumber\\
&&\mbox{}\times(\tau_1+\tau_2)
\delta\left(\sigma-(\Delta/2\pi\hbar)c(\tau_1-\tau_2) \sqrt{1/T-1}\right).
\end{eqnarray}
As in Refs.\ \onlinecite{brouwer} and \onlinecite{pedersen}, the density of states
appears as a weight factor $\tau_1+\tau_2$ in the ensemble average
(\ref{tpdistcav}), because the ensemble is generated by uniformly varying the charge
on the quantum dot rather than its Fermi energy. This is the correct thing to do in
the Hartree (self-consistent potential) approximation. A more sophisticated treatment of
the electron-electron interactions (as advocated in Ref.\ \onlinecite{aleiner}) does not
yet exist for this problem. The resulting distributions
are plotted in Fig.\ \ref{figcav}.  The curves have a cusp at $\sigma=0$, and
asymptotes $P(\sigma)\propto |\sigma|^{-1-\beta}\ln {|\sigma|}$ for $|\sigma|\gg1$.

\section{Conclusion}
The results we have reported hold for single-channel conductors.
The generalization to multi-channel conductors is of interest. Multi-channel
diffusive conductors were studied in Ref.\ \onlinecite{diffusive}. For a chaotic 
cavity with ballistic point contacts having a large number of modes
($N$ modes per point contact), the distribution of the thermopower is Gaussian.
The mean is zero and the variance is
\begin{equation}
{\rm Var}\, S = \frac{k_{\rm B}^4T^2\pi^6}{9e^2N^4\Delta^2\beta}.
\end{equation}
(We have used the results of Ref.\ \onlinecite{efetov}.) Analogously to universal
conductance fluctuations, the variance of the thermopower is reduced by a factor of
$2$ upon breaking time-reversal symmetry ($\beta=1\rightarrow\beta=2$).

For an $N$-mode wire in the localized regime, our derivation of the exponential
distribution of the thermopower remains valid. This is not true for the
Lorentzian distribution. The reason is that the Herbert-Jones-Thouless formula
for $N>1$ relates the density of states to the sum of the inverse localization lengths,
\cite{craig}
and there is no simple relation between this sum and the thermopower.
We expect that the tail of the distribution remains quadratic,
$P(S)\propto S^{-2}$ --- because of the argument of Sec.\ \ref{fintemp}, which
is still valid for $N>1$. It remains a challenge to determine analytically the entire
thermopower distribution of a multi-channel disordered wire.

\acknowledgements
This paper is dedicated to Rolf Landauer on the occasion of his 70th birthday.
Discussions with P. W. Brouwer are gratefully acknowledged.
This research was supported by the ``Ne\-der\-land\-se or\-ga\-ni\-sa\-tie voor
We\-ten\-schap\-pe\-lijk On\-der\-zoek'' (NWO) and by the ``Stich\-ting voor
Fun\-da\-men\-teel On\-der\-zoek der Ma\-te\-rie'' (FOM).

\end{document}